\documentclass[aip,pof,reprint,onecolumn]{revtex4-1}
\usepackage{amsmath}
\usepackage{graphicx}
\usepackage{color}
\usepackage{hyperref}
\usepackage{psfrag}
\usepackage{stmaryrd}
\usepackage{amssymb}
\usepackage{wasysym}

\def \mps {m~s$^{-1}$}

\usepackage{geometry}
\newcommand{\modif}[1]{\textcolor{black}{#1}}

\begin{document}

\title{Surface deformations and wave generation by wind \modif{blowing over} a viscous liquid}
\author{A. Paquier}
\author{F. Moisy}
\author{M. Rabaud}
\affiliation{Laboratoire FAST, Univ. Paris-Sud, CNRS, Universit\'e Paris-Saclay, F-91405, Orsay, France.}

%\modif{VERSION 3}
\date{\today}

%\pacs{47.35.-i}{Hydrodynamic waves}
%\pacs{47.54.-r}{Pattern selection; pattern formation (fluid dynamics)}

\begin{abstract}

We investigate experimentally the early stage of the generation of waves by a turbulent wind at the surface of a viscous liquid. The spatio-temporal structure of the surface deformation is analyzed by the optical method Free Surface Synthetic Schlieren, which allows for time-resolved measurements with a micrometric accuracy. Because of the high viscosity of the liquid, the flow induced by the turbulent wind in the liquid remains laminar, with weak surface drift velocity.  Two regimes of deformation of the liquid-air interface are identified. In the first regime, at low wind speed, the surface is dominated by rapidly propagating disorganized wrinkles, elongated in the streamwise direction, which can be interpreted as the surface response to the pressure fluctuations advected by the turbulent airflow. The amplitude of these deformations increases approximately linearly with wind velocity and are essentially independent of the fetch (distance along the channel). Above a threshold in wind speed, the perturbations organize themselves spatially into quasi parallel waves perpendicular to the wind direction with their amplitude increasing downstream. In this second regime, the wave amplitude increases with wind speed but far more quickly than in the first regime.

\end{abstract}

\maketitle

\section{Introduction} \label{sec_intro}

Understanding the generation of surface waves under the action of wind is an old problem which is of primary interest for wave forecasting and to evaluate air-sea exchanges of heat, mass and momentum \modif{on Earth\cite{leblond1981waves, Janssen_2004} or on natural satellites.\cite{Hayes_2013, barnes2014cassini}} It is also important in engineering applications involving liquid and gas transport in pipes.\cite{hewitt2013annular} Despite the considerable literature on the subject, the physical mechanism for the onset of the first ripples at low wind velocity is still not fully understood. \modif{Russell,\cite{russell1844waves} as quoted by Kelvin,\cite{thomson1887waves} nicely described the first regime where a very slight wind first destroys the perfect mirror reflection of the water surface, followed by a second regime for slightly larger wind where waves are observed.}  The first attempt to explain the wind-wave formation was proposed by Helmholtz and Kelvin,\cite{Helmholtz_1868, thomson1887waves} and the Kelvin-Helmholtz instability is now a paradigm for instabilities in fluid mechanics. However, Kelvin was aware of the discrepancy between the predicted critical wind of 6.6~\mps~ and the commonly observed minimal wind of the order of 1~\mps~ for the first visible ripples on a calm sea.\cite{Darrigol} He ascribed this discrepancy to viscous effects, which were not taken into account in the model. Since then, numerous attempts to better predict the onset of wind waves were proposed, still with limited success.

Among the large body of literature on the subject, pioneering theoretical contributions are those of Phillips\cite{phillips1957generation} and Miles.\cite{Miles_1957} In an enlightening paper, Phillips\cite{phillips1957generation} analyzed how pressure fluctuations in the turbulent air boundary layer could deform an otherwise inviscid fluid at rest. He suggested that the pressure perturbations whose size and phase velocity match that of the waves are selectively amplified by a resonance mechanism, and obtained a linear growth in time of the squared wave amplitude. The same year, Miles\cite{Miles_1957} proposed another mechanism based on the shear flow instability of the mean air velocity profile, ignoring viscosity, surface tension, drift of the liquid and turbulent fluctuations. From a temporal stability analysis, he showed that the boundary layer in the air is unstable if the curvature of the velocity profile is negative at the critical height at which air moves at the phase velocity of the waves, resulting in an exponential growth in time of the wave amplitude. An effort to classify the various instability mechanisms in parallel two-phase flow, including Miles', is proposed in the review by Boomkamp and Miesen.\cite{boomkamp1996classification}

Since then, many attempts have been made to test these predictions\cite{plate1969experiments, Kahma_1988, Liberzon_2011, grare2013growth, hristov2003dynamical} or to improve these models,\cite{katsis1985wind, teixeira2006initiation, valenzuela1976growth, Gastel1985phase, young2014generation} with no definitive conclusion at the moment. While several experiments were devoted to determine the temporal growth of the wave after a rapid initiation of the wind,\cite{mitsuyasu1978growth, Kawai1979generation, veron2001experiments} other tested the amplification by wind of mechanically generated waves\cite{bole1969response,gottifredi1970growth, Wilson_1973, mitsuyasu1982wind,tsai2005spatial} or the wave formation by a laminar air flow.\cite{tsai2005spatial, Gondret97b, Gondret99} Since the boundary layers in both fluids are generally turbulent in the case of the air-water interface,\cite{hidy1966wind, Caulliez_1998, caulliez2007turbulence, longo2012_turbulent} some authors simplified the problem by considering more viscous liquids.\cite{keulegan1951wind, Francis_1954, gottifredi1970growth, Gondret97b, naraigh2011interfacial} With an airflow above a liquid more viscous than water, the wave onset is larger and, paradoxically, in better agreement with the inviscid Kelvin-Helmholtz prediction.\cite{Francis_1954, Miles1959generation, Gondret97b}

Rapid progresses in numerical simulations have made it possible now to address the coupled turbulent flows of air and water and their effect on the interface, and to access the pressure and stress fields hardly measurable in experiments.\cite{lin2008direct} On the experimental side, recent improvements in optical methods have opened the possibility to access experimentally the spatio-temporal structures of the waves with unprecedented resolution.\cite{Moisy09, kiefhaber2014high} 

 In the present work, we take advantage of this technical improvement to analyze the early stage of wave formation at the surface of a viscous liquid. Surface deformations are measured with a vertical resolution better than one micrometer using Free-surface Synthetic Schlieren,\cite{Moisy09} a time-resolved optical method based on the refraction of a pattern located below the fluid interface.
Working with a viscous liquid has two advantages: first, the flow in the liquid remains laminar and essentially unidirectional with a limited surface drift; second, the perturbations of the interface that are not amplified by an instability mechanism are rapidly damped, so the surface deformations at low wind velocity are expected to be the local response in space and time to the instantaneous pressure fluctuations in the air. Our results clearly exhibit two wave regimes: (i) at low wind velocity, small disordered surface deformations that we call "wrinkles" first appear, elongated in the streamwise direction, with amplitude growing slowly with the wind velocity but with no significant evolution with fetch (the distance upon which the air blows on the liquid); (ii) above a well defined wind velocity, a regular pattern of gravity-capillary waves appears, with crests normal to the wind direction and amplitude rapidly increasing  with wind velocity and fetch.

\section{Experimental set-up}
\label{sec:experimental}

\subsection{Liquid tank and wind tunnel}

\begin{figure}
\centerline{\includegraphics[width=14cm]{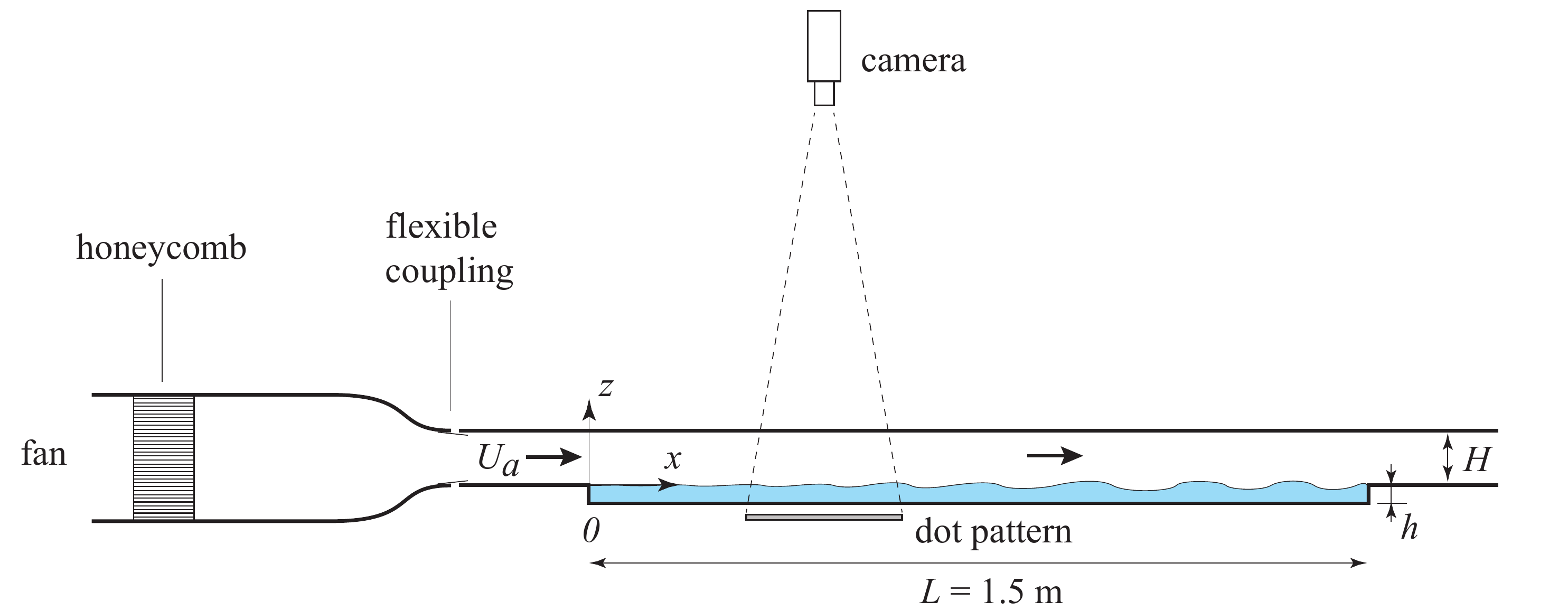}}
\caption{Experimental setup. The wave tank and the wind tunnel are connected to the upstream air flow via a flexible coupling to minimize transmission of vibrations induced by the centrifugal fan. The surface deformations are measured by Free-Surface Synthetic Schlieren, by imaging from above a pattern of random dots located below the liquid tank. }
\label{fig:Wind_tunnel}
\end{figure}

The experimental set-up is sketched in Fig.~\ref{fig:Wind_tunnel}. It is composed of a fully transparent Plexiglas rectangular tank of length $L=1.5$~m, width $W=296$~mm, and depth $h=35$~mm, fitted to the bottom of a horizontal channel of rectangular cross-section. The channel width is identical to that of the tank, its height is $H=105$~mm, with two horizontal floors of length 26~cm before and after the tank. The tank is filled with a water-glycerol mixture, such that the surface of the liquid precisely coincides with the bottom of the wind tunnel.

%%%%%%%%%%%%%%
%%%%%%%%%%%%%%

Air is injected upstream by a centrifugal fan through a honeycomb and a convergent (ratio 2.4 in the vertical direction). To minimize transmission of vibrations induced by the fan, the wind-tunnel is mounted on a heavy granite table and connected to the upstream channel via a flexible coupling. \modif{Residual vibrations induce surface deformations less than 1~$\mu$m.} The wind velocity $U_a$, measured at the center of the outlet of the wind tunnel with a hot-wire anemometer, can be adjusted in the range $1-10$~\mps.  We define $x$ in the streamwise direction (fetch), $y$ in the spanwise direction and $z$ in the vertical direction. The origin (0,0,0) is located at the free surface at fetch 0, at mid-distance between the lateral walls.

The tank is filled with a mixture of 80\% glycerol and 20\% water, of density $\rho = 1.20 \times 10^3 $~kg~m$^{-3}$ at 25$^\mathrm{o}$C (the room temperature being regulated to this temperature). Kinematic viscosity, measured with a low shear rheometer, is $\nu =\eta / \rho = 30 \times 10^{-6}$~m$^2$~s$^{-1}$ at this temperature. The water-glycerol mixture is extremely sensitive to surface contamination, which may induce strong surface tension gradients and alter both the mean flow in the liquid and the generation of waves.\cite{Kahma_1988} To overcome this problem, we let the wind blow for a few minutes, and we remove the contaminated part of the surface liquid by collecting it at the end of the tank. The procedure is repeated frequently, and in normal operating conditions the surface of the liquid remains clean over most of the liquid bath, with less than 30~cm of polluted surface remaining at the end of the tank. Surface tension of the clean mixture, measured with a Wilhelmy plate tensiometer, is $\gamma = 60 \pm 5$ mN~m$^{-1}$, and the capillary wavelength is $\lambda_c=2\pi \sqrt{\gamma / \rho g} \simeq 14.2$~mm. The  dispersion relation for free surface waves propagating in \modif{an inviscid} liquid at rest is
\begin{equation}
\omega^2=\left(gk+\frac{\gamma}{\rho} k^3\right)\tanh(kh),
\label{eq:dispersion_relation}
\end{equation} 
where $\omega$ and $k$ are the angular frequency and wave number. Finite depth effects can be neglected in the present experiment: the depth correction factor, $\tanh(kh)$, is larger than 0.98 for wavelength smaller than 90~mm.
\modif{In spite of the large viscosity used in our experiments, viscous correction to the inviscid dispersion relation can be also neglected here:\cite{Lamb,padrino2007correction} the phase velocity matches the inviscid prediction to better than $10^{-3}$ for the waves observed at onset ($\lambda \simeq 30$ mm). On the other hand, this large viscosity induces a strong attenuation of the waves. For the wave tank geometry and the typical wavelengths considered here, friction with the bottom and side walls is negligible, and the attenuation length for free waves is governed by the dissipation in the bulk,\cite{Lighthill} $L_v = c_g / (2 \nu k^2)$, with $c_g(k)$ the group velocity. For $\lambda \simeq 30$ mm the attenuation length is $L_v \simeq 60$~mm, indicating that a free disturbance at this \modif{wavelength} cannot propagate over a distance much larger than a few wavelengths.} As a consequence, although the tank is of limited size, reflections on the walls or at the end of the tank can be neglected in our experiment.

\subsection{Wind profile}

%%%%%%%%%%%%%%
\begin{figure}
\centerline{\includegraphics[width=13cm]{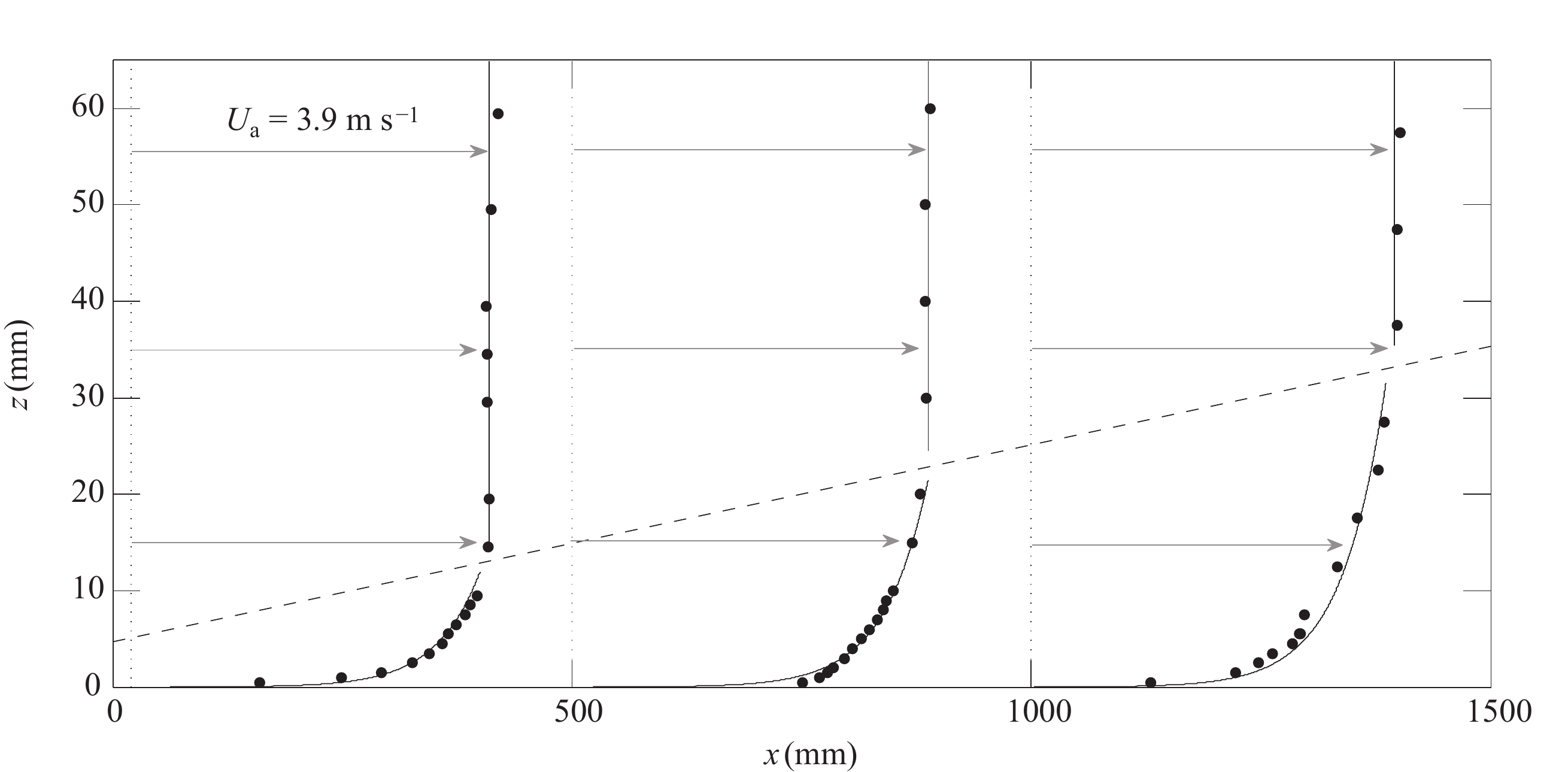}}
\caption{Mean velocity profiles $U(z)$ in the air for $U_a = 3.9$~\mps~ at fetch $x = 20$, 500 and 1000~mm and $y=0$. The dotted line shows the streamwise development of the 99\% boundary-layer thickness, $\delta_{0.99}(x) \simeq 12.6$~mm + $0.02 x$. The curves for $z < \delta_{0.99}(x)$ show the fit with the logarithmic law (\ref{eq:CL_Turbulente}).}
\label{fig:airprof}
\end{figure}
%%%%%%%%%%%%%%

The velocity profile in the air $U(z)$, measured using hot-wire anemometry, is shown in Fig.~\ref{fig:airprof} for a wind velocity $U_a = 3.9$~\mps~ at fetch $x=20$, $500$ et $1000$ mm. The hot-wire (Dantec Dynamics 55P01) is 5~$\mu$m in diameter with an active length of 1.25~mm, and is mounted on a sliding arm to allow vertical motion with a 0.1~mm accuracy. The velocity profiles show the development of the boundary layer along the channel:
the thickness $\delta_{0.99}$, defined as the distance from the surface at which the mean velocity is $0.99 U_a$, increases nearly linearly, from 12.6~mm at $x=0$ to 32~mm at $x = 1.0$~m (slope of order of 2\%). The fact that $\delta_{0.99}(x)$ approaches the channel half-height $H/2 \simeq 52$~mm at the end of the channel indicates that the flow becomes fully developed there.

The evolution of the friction velocity $u^*(x)$ along the channel can be obtained by fitting the velocity profiles for $z < \delta_{0.99}(x)$
with the classical logarithmic law,\cite{Schlichtling, plate1969experiments}
\begin{equation}
\frac{U(x,z)}{u^*(x)} = \frac 1 \kappa \ln \left(\frac {z}{\delta_v(x)}\right) + C,
\label{eq:CL_Turbulente}
\end{equation}
with $\kappa \simeq 0.4$ the K\'arm\'an constant, $C=5$, and $\delta_v(x) = \nu_a /u^*(x)$ the thickness of the viscous sublayer.
We find $u^*$ to slightly decrease with fetch: for $U_a = 3.9$~\mps, $u^*$ decreases from 0.22~\mps~ at $x \simeq 0$ down to 0.17~\mps~ at $x = 1$~m. Accordingly, $\delta_v(x)$ slightly increases with fetch, from 0.07 to 0.09~mm.

The procedure is repeated for different wind velocities at a fixed fetch, $x_0 = 500$~mm. Measurements are restricted to $U_a < 6$~\mps, when the surface deformations remain weak (less than 10~$\mu$m), because the hot-wire could not be positioned too close to the liquid. We find that in this range $u^*$ is almost proportional to $U_a$ , $u^*(x_0) \simeq 0.05 U_a$ (see inset in Fig.~\ref{fig:PIV_profiles}). The corresponding half-height channel Reynolds number at this fetch,  $Re_\tau = H u^* / 2\nu_a$, varies in the range $160-1000$, and the thickness of the viscous sublayer $\delta_v$ decreases from 0.3 to 0.05~mm when $U_a$ increases from 1 to 6~\mps. Since the flow in the viscous sublayer is essentially laminar up to $z \simeq 10 \delta_v$, which is comfortably larger than any surface deformation over this range of velocity, we can consider the air flow to be close to a canonical turbulent boundary over a no-slip flat wall, at least for a wind velocity up to 6~\mps.  \modif{This does not hold for larger wind velocity, for which the roughness induced by the waves decreases the value of $C$ in Eq.~(\ref{eq:CL_Turbulente}).\cite{longo2012wind, zavadsky2012characterization}}

\subsection{Flow in the liquid tank}

%%%%%%%%%%%%%%
\begin{figure}
\centerline{\includegraphics[width=11cm]{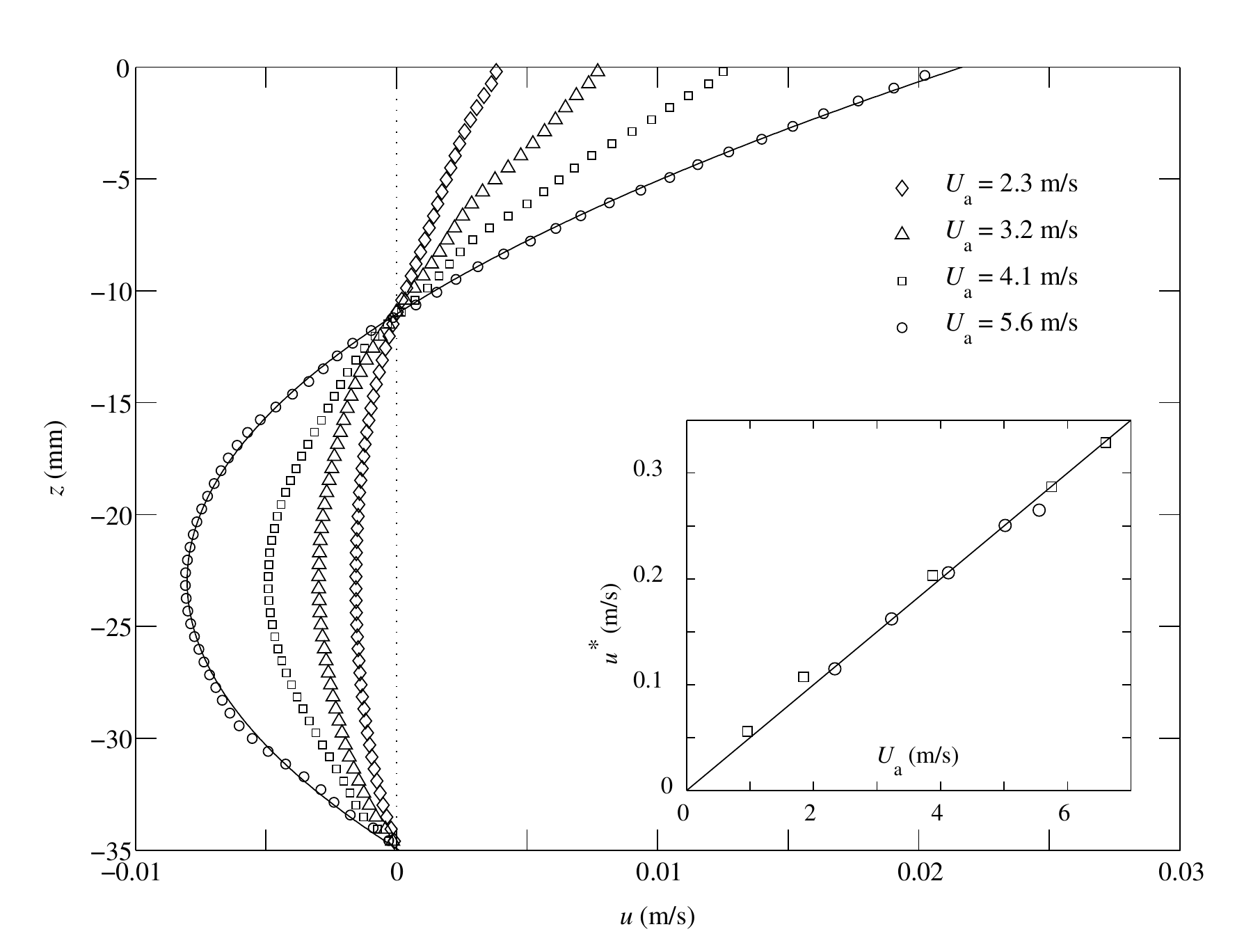}}
\caption{Mean velocity profiles in the liquid measured by PIV at fetch $x=400$ mm for various wind velocities $U_a$. The profiles are averaged in time and in the streamwise direction over $\Delta x =100$~mm. The continuous line for the largest value of $U_a$ shows the quadratic profile (\ref{eq:parabolic_profil}). Inset: friction velocity $u^*$, deduced from the mean profile in the airflow at $x_0 =500$~mm (squares) and deduced from the shear stress at the liquid surface (Eq. \ref{eq:u_star}) (circles), as a function of the wind velocity $U_a$. The continuous line is a fit by $u^*=0.05 U_a$.}
\label{fig:PIV_profiles}
\end{figure}
%%%%%%%%%%%%%%

The shear stress induced by the wind at the interface drives a drift flow in the liquid. Since the tank is closed, this drift is compensated by a back-flow at the bottom of the tank, and a stationary state is reached after a few minutes. We have measured the mean velocity profile in the tank using Particule Image Velocity (PIV) in vertical planes $(x,z)$ (Fig.~\ref{fig:PIV_profiles}). Except at small fetch (on a distance of the order of the liquid height) and over the last 30~cm of the tank (where surface contamination cannot be avoided), the velocity profiles are found nearly homogeneous in $x$ and $y$. The velocity profiles are well described by the parabolic law, solution of the stationary Stokes problem
\begin{equation}
u(x,z)=U_s(x) \left( 1+\frac{z}{h} \right) \left(1+3 \frac{z}{h} \right),
\label{eq:parabolic_profil}
\end{equation}
for $-h \leq z \leq 0$, where $U_s(x) = u(x,z=0)$ is the surface velocity. For $U_a = 4$~\mps
the surface velocity is of order of 1~cm~s$^{-1}$, \modif{which leads to a Reynolds number $Re = U_s h /\nu \simeq 10$ in the liquid. This drift velocity is in agreement with measurements at small Reynolds number,\cite{keulegan1951wind} but is much smaller than the 2-3\% of wind velocity typically found in classical air-water experiments.\cite{plate1969experiments, gottifredi1970growth, tsai2005spatial, Liberzon_2011}} The small surface velocity here is expected to have negligible effect on the dispersion relation (\ref{eq:dispersion_relation}): Lilly (see appendix of Hidy and Plate\cite{hidy1966wind}) shows that the correction to the phase velocity for this parabolic profile is $2U_s / kh$, which is only 2\% of the phase velocity for the most unstable wavelength ($\lambda \simeq 30$~mm).

Because of the development of the boundary layer and the resulting decreasing friction velocity, the surface velocity $U_s$ decreases slightly along the tank. For $U_a = 4$~\mps, $U_s(x)$ decreases from 1.5 to 1.1~cm~s$^{-1}$.  Measuring $U_s(x)$ provides another way to determine $u^*(x)$: using the continuity of the stress at the interface, one has $\sigma(x) = \rho_a u^{*2}(x) = \eta \partial u / \partial z (z=0)$, yielding
\begin{equation}
u^*=\sqrt{\frac{4\eta U_s}{\rho_a h}}.
\label{eq:u_star}
\end{equation}
The friction velocity $u^*$ measured by PIV in the liquid with this method is in excellent agreement with the one measured in air with the hot-wire at $x_0 =500$~mm (inset of Fig.~\ref{fig:PIV_profiles}). For simplicity, the ratio of $u^*/U_a$ is taken in the following as constant and equal to 0.05 for all fetches. By comparison, the ratio of $u^*/U_a$ is generally found of order of 3\% in air-water experiments,\cite{plate1967laboratory, mitsuyasu1978growth, wu1975wind} with weak dependence on the wind velocity.\cite{mitsuyasu1978growth}

\modif{Note that the shear stress at the liquid surface and at the lateral and upper walls must be balanced by a small longitudinal pressure gradient $\Delta p / L$ in the air along the channel. This pressure gradient introduces a complication in the setup: the liquid surface becomes slightly tilted, with the inlet liquid height below the outlet height (this is analogous to the 'wind tide' effect observed on lakes\cite{keulegan1951wind}). Assuming equal stress $\sigma$ on the liquid surface and on the solid walls}, this pressure gradient writes $\Delta p / L \simeq 2 \sigma (1/W + 1/H)$, with $W$ and $H$ the channel width and height. For a wind velocity $U_a = 4$~\mps, the pressure drop along the tank is $\Delta p \simeq 2$~Pa, which results in a hydrostatic height difference between the two ends of the tank of $\Delta p / \rho g \simeq 0.2$~mm, in good agreement with our measurement. We observed that the resulting backward facing step that could appear at $x = 0$ increases the turbulent fluctuations and significantly enhances the wave amplitude at small fetch by typically a factor of 2. It is therefore critical to maintain the liquid level at $x=0$ by carefully tilting the channel. We achieve a leveling of the liquid at $x=0$ better than 20 $\mu$m by using the tangential reflexion of a laser sheet intersecting the upstream plate and the liquid surface.

\subsection{Surface deformation measurement}

We measure the surface deformation of the liquid using the Free Surface Synthetic Schlieren (FS-SS) method.\cite{Moisy09} This optical method 
is based on the analysis of the refracted image of a pattern visualized through the interface. A random dot pattern located below the liquid tank is imaged by a fast camera located above the channel, with a field of view of $390 \times 280$~mm.
A reference image is taken when the liquid surface is flat (zero wind), and the apparent displacement field $\delta {\bf r}$ between this 
reference image and the distorted image in the presence of waves is computed using an image correlation algorithm. Integration of this displacement field gives the height field $\zeta(x,y,t)$ (see examples in Fig.~\ref{fig:snapshots}).

Measurements are performed at three fetches, corresponding to the the first three quarters of the tank with a small overlap: $x \in [10, 400]$~mm; $x \in [370, 760]$~mm; $x \in [700, 1090]$~mm. No measurements are performed in the last quarter of the channel because of possible surface contamination.
The distance between the random dot pattern and the liquid surface sets the sensitivity of the measurement, and is chosen according to the typical wave amplitude. We chose a distance of 29~cm for waves of weak amplitude (of order of $1-10~\mu$m), and
6~cm for waves of large amplitude (up to 1~mm). For wave amplitude larger than a few millimeters, the FF-SS method no longer applies: crossing of light rays appear below waves of large curvature (caustics), which prevents the measurement of the apparent displacement field. The horizontal resolution is 3~mm, and the vertical resolution of order of 1\% of the wave amplitude. Acquisitions of 2~s at 200~Hz are performed for time-resolved wave reconstruction, and 100~s at 10~Hz to ensure good statistical convergence of the root mean square of the wave amplitude.

\section{Results}
\label{sec:deformations}

\subsection{Amplitude versus velocity}

%%%%%%%%%%%%%%
\begin{figure}
\centerline{\includegraphics[width=14cm]{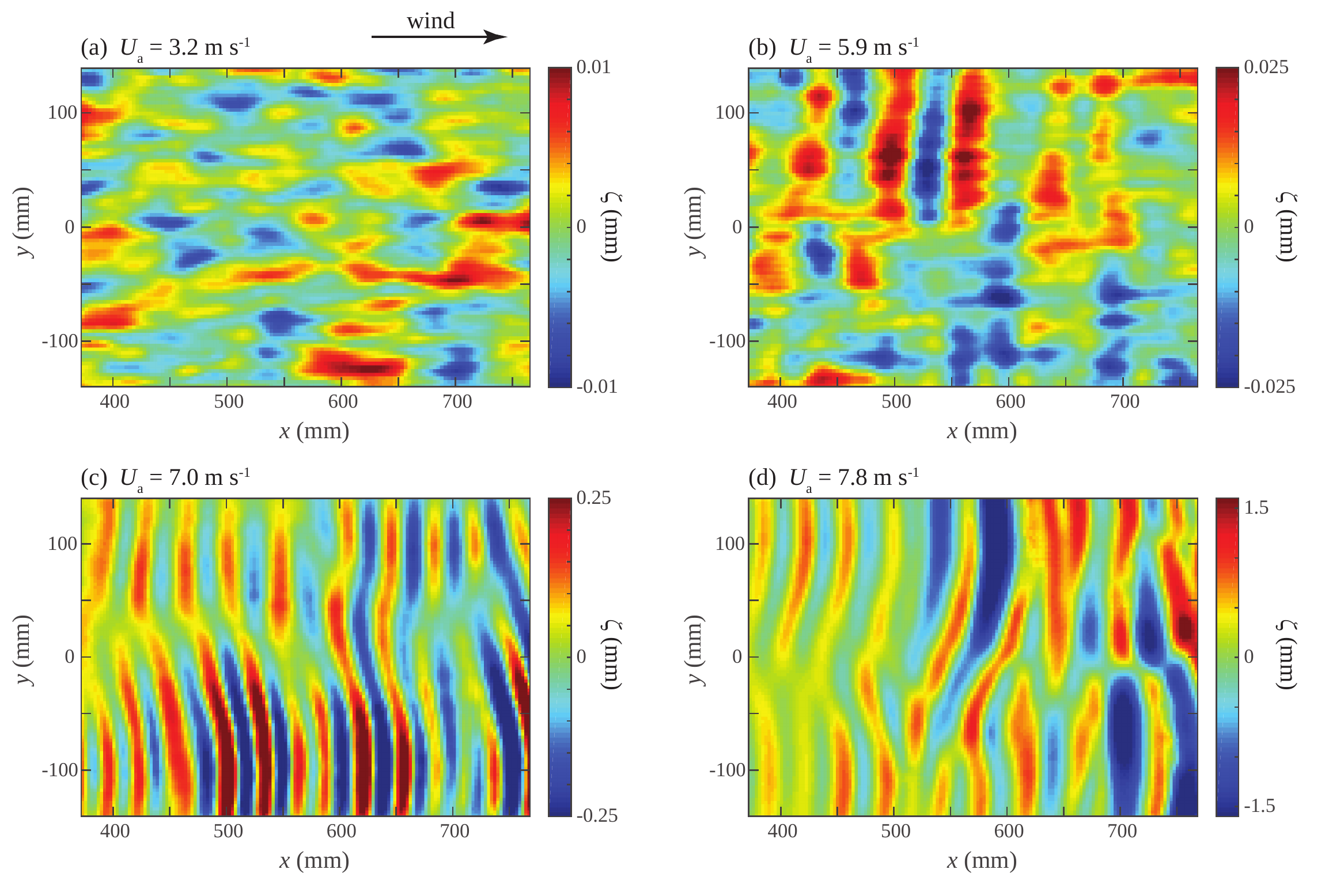}}
\caption{Instantaneous surface height $\zeta(x,y)$ measured by FS-SS centered at
intermediate fetch $x= 570$~mm, at increasing wind velocities. (a) $U_a=3.2$~\mps, showing small-amplitude disorganized wrinkles elongated in the streamwise direction ($\zeta_{\rm rms}= 0.0032$~mm). (b) $U_a=5.9$~\mps, showing a combination of streamwise wrinkles and spanwise waves ($\zeta_{\rm rms}= 0.009$~mm). (c) $U_a=7.0$~\mps, showing well-defined spanwise waves of mean wavelength $\lambda=35$~mm ($\zeta_{\rm rms}= 0.12$~mm). (d) $U_a = 7.8$~\mps, showing large-amplitude waves of mean wavelength $\lambda=44$~mm with increasing disorder ($\zeta_{\rm rms}= 0.6$~mm). Note the change of scale in the color map.}
\label{fig:snapshots}
\end{figure}
%%%%%%%%%%%%%%

Figure~\ref{fig:snapshots} shows four snapshots of the surface deformation at increasing wind velocity, between $U_a = 3.2$ and 7.8~\mps, at intermediate fetch $x \in [370, 760]$~mm. At small $U_a$ the wave pattern shows rapidly moving disorganized wrinkles of weak amplitude, of order 10~$\mu$m, elongated in the streamwise direction (Fig.~\ref{fig:snapshots}a). As the wind velocity is increased, noisy spanwise crests, normal to the wind direction, gradually appear in addition to the streamwise wrinkles (Fig.~\ref{fig:snapshots}b). The amplitude of these spanwise crests rapidly increases with velocity and becomes much larger than the amplitude of the streamwise wrinkles for wind velocity in the range $6-7$~\mps. At $U_a = 7.0$~\mps (Fig.~\ref{fig:snapshots}c) the surface field is dominated by a regular wave pattern of typical amplitude 0.2~mm, with a well defined wavelength in the streamwise direction. The wave crests are not strictly normal to the wind, but rather show a dislocation near the center line. \modif{This may be due to a slight convergence of the turbulent air flow close to the free surface towards the walls, which is unavoidable for a turbulent channel flow in a rectangular geometry (secondary flow of Prandtl's second kind\cite{Schlichtling})}. As the wind speed is further increased, the disorder of the wave pattern increases (Fig.~\ref{fig:snapshots}d), with more dislocations and larger typical wavelength and amplitude. All these patterns are quite similar to those reported by Lin {\it et al.}\cite{lin2008direct} from direct numerical simulation of temporally growing waves with periodic boundary conditions.

%%%%%%%%%%%%%%
\begin{figure}
\centerline{\includegraphics[width=10cm]{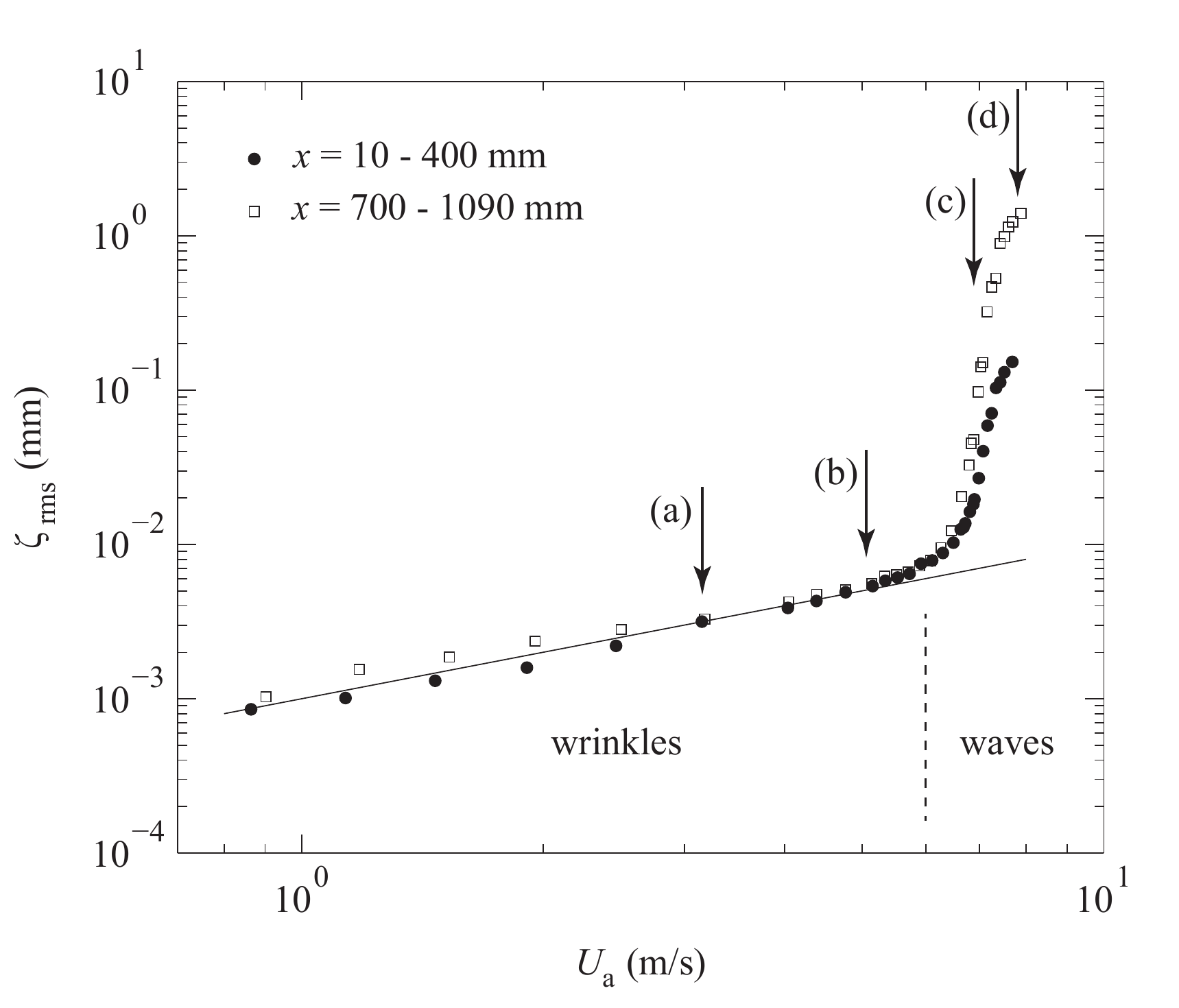}}
\caption{Root mean square of the surface height $\zeta_{\rm rms}$ as a function of wind velocity $U_a$. The data are averaged over the measurement windows centered at two values of the fetch $x$. The vertical arrows show the velocities corresponding to the four snapshots in Fig.~\ref{fig:snapshots}. The continuous line shows the linear fit $\zeta_{\rm rms} = \alpha U_a$, with $\alpha = 10^{-6}$~s.}
\label{fig:rms}
\end{figure}
%%%%%%%%%%%%%%

The evolution from the disorganized longitudinal wrinkles to the well-defined transverse waves as the wind velocity is increased is evident from the root mean square of the deformation amplitude,
$$
\zeta_{\rm rms} = \langle \zeta^2(x,y,t) \rangle^{1/2},
$$
where the brackets are both temporal average and spatial average over the field of view. This quantity, plotted as a function of the wind velocity in Fig.~\ref{fig:rms} for two values of the fetch $x$, clearly exhibits the two regimes: at small air velocity, when the surface deformation is dominated by the longitudinal wrinkles, the wave height slowly increases with the wind velocity, but beyond a threshold of order of 6~\mps~ the increase becomes much sharper and fetch dependent: the wave amplitude grows by a factor of 100 for $U_a$ increasing between 6 and 8~\mps. \modif{A similar transition in the wave amplitude is also reported in air-water experiments by Kahma and Donelan\cite{Kahma_1988} and Caulliez et al.\cite{Caulliez_1998}} At the largest velocity, $U_a \simeq 8$~\mps, the sharp increase of the wave amplitude apparently starts to saturate. This wind velocity represents an upper limit for the FS-SS measurements because of the caustics induced by the strong wave curvature. 

In the wrinkle regime ($U_a < 6$~\mps), the wave height is almost independent of the fetch, and is approximately proportional to the wind velocity: $\zeta_{\rm rms} \simeq \alpha U_a$, with $\alpha = 10^{-6}$~s. This independence of $x$ suggests that the wrinkles can be simply viewed as an imprint on the free surface of the turbulent fluctuations in the airflow. Relating quantitatively the height fluctuations to the pressure fluctuations is however a difficult task. A simple estimate, assuming an instantaneous hydrostatic response of the liquid interface (i.e., neglecting viscous and capillary effects) would yield $\zeta_{\rm rms} \simeq p_{\rm rms} / \rho g$. The pressure fluctuation at the wall in a fully developed turbulent channel is well described by the empirical law\cite{hu2006wall, jimenez2008turbulent} $p_{\rm rms} = f(Re_\tau) \rho_a u^{*2}$, with $f(Re_\tau) = (2.60 \ln (Re_\tau) - 11.25)^{1/2}$. In the range $U_a \simeq 1-6$~\mps, one has $Re_\tau \simeq 160-1000$, which (neglecting the logarithmic variation over this range and taking $u^* \simeq 0.05 U_a$) yields $p_{\rm rms} \simeq 0.006 \rho_a U_a^2$, and hence $\zeta_{\rm rms} \simeq 0.3-20$~$\mu$m. Although the order of magnitude is consistent with Fig.~\ref{fig:rms}, the predicted scaling ($\zeta_{\rm rms} \propto U_a^2$) is not compatible with the data, \modif{suggesting that the viscous time response of the liquid must be accounted for to describe the observed trend $\zeta_{\rm rms} \propto U_a$.}

%%%%%%%%%%%%%%
\begin{figure}
\centerline{\includegraphics[width=12cm]{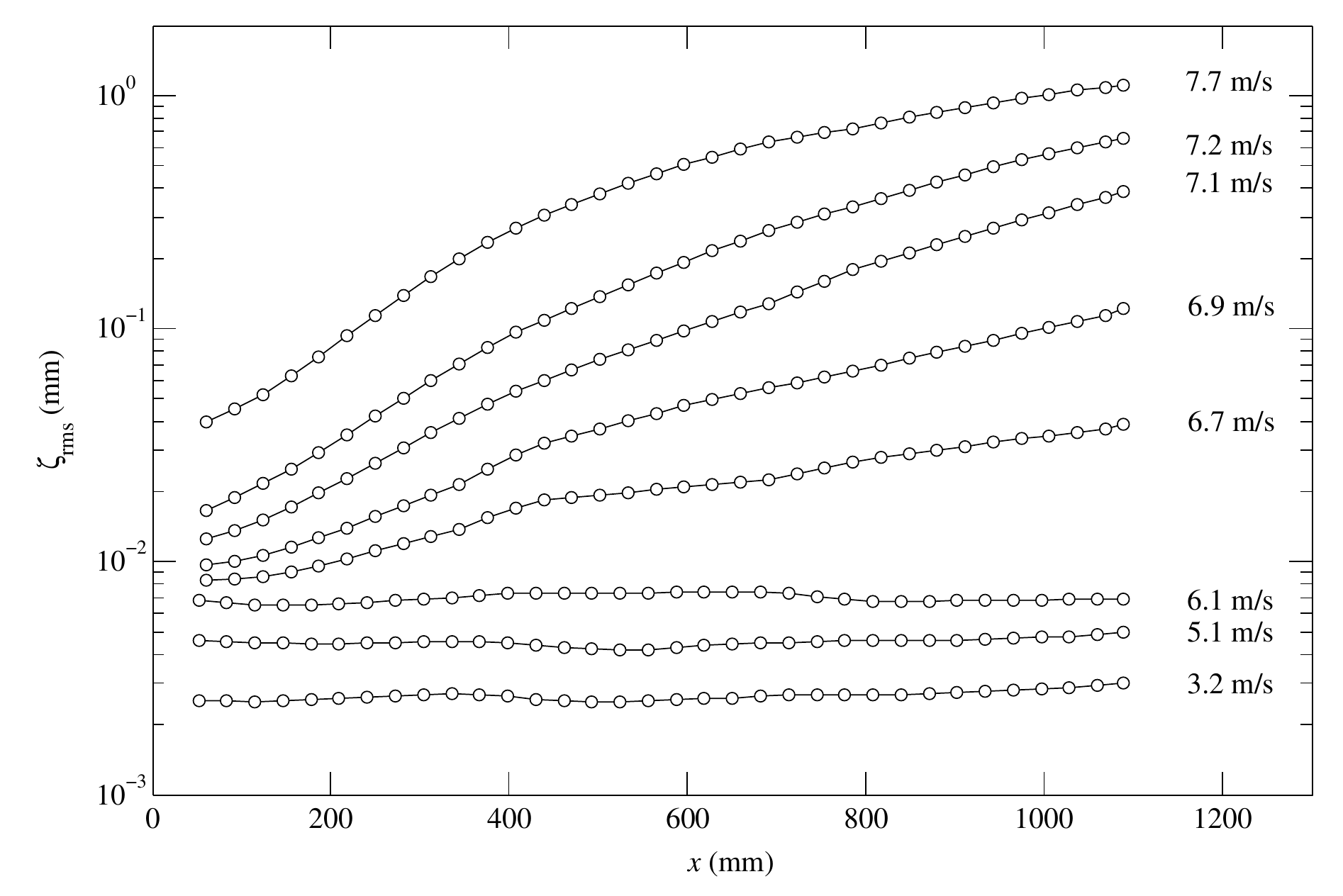}}
\caption{Wave amplitude $\zeta_{\rm rms}$ (averaged over the spanwise coordinate $y$ and time) as a function of the fetch $x$ for various wind speeds $U_a$.  In the wrinkle regime ($U_a < 6.3$~\mps) $\zeta_{\rm rms}(x)$ is almost constant, whereas it increases with fetch in the wave regime.}
\label{fig:rmsvsfetch}
\end{figure}
%%%%%%%%%%%%%%

\subsection{Spatial growth rate}

Contrarily to the wrinkles, which are almost independent of the fetch $x$, the amplitude of the transverse waves strongly increases with $x$, as shown in Fig.~\ref{fig:rmsvsfetch}. The rms amplitude $\zeta_{\rm rms}(x)$ is computed here using an average over $y$ and time only. \modif{The spatial growth is approximately exponential at small fetch ($x < 400$~mm), as expected for a convective supercritical instability in an open flow.\cite{huerre1998hydrodynamic} At larger fetch nonlinear effects come into play, resulting in a weaker growth of the wave amplitude.}

\begin{figure}
\centerline{\includegraphics[width=9cm]{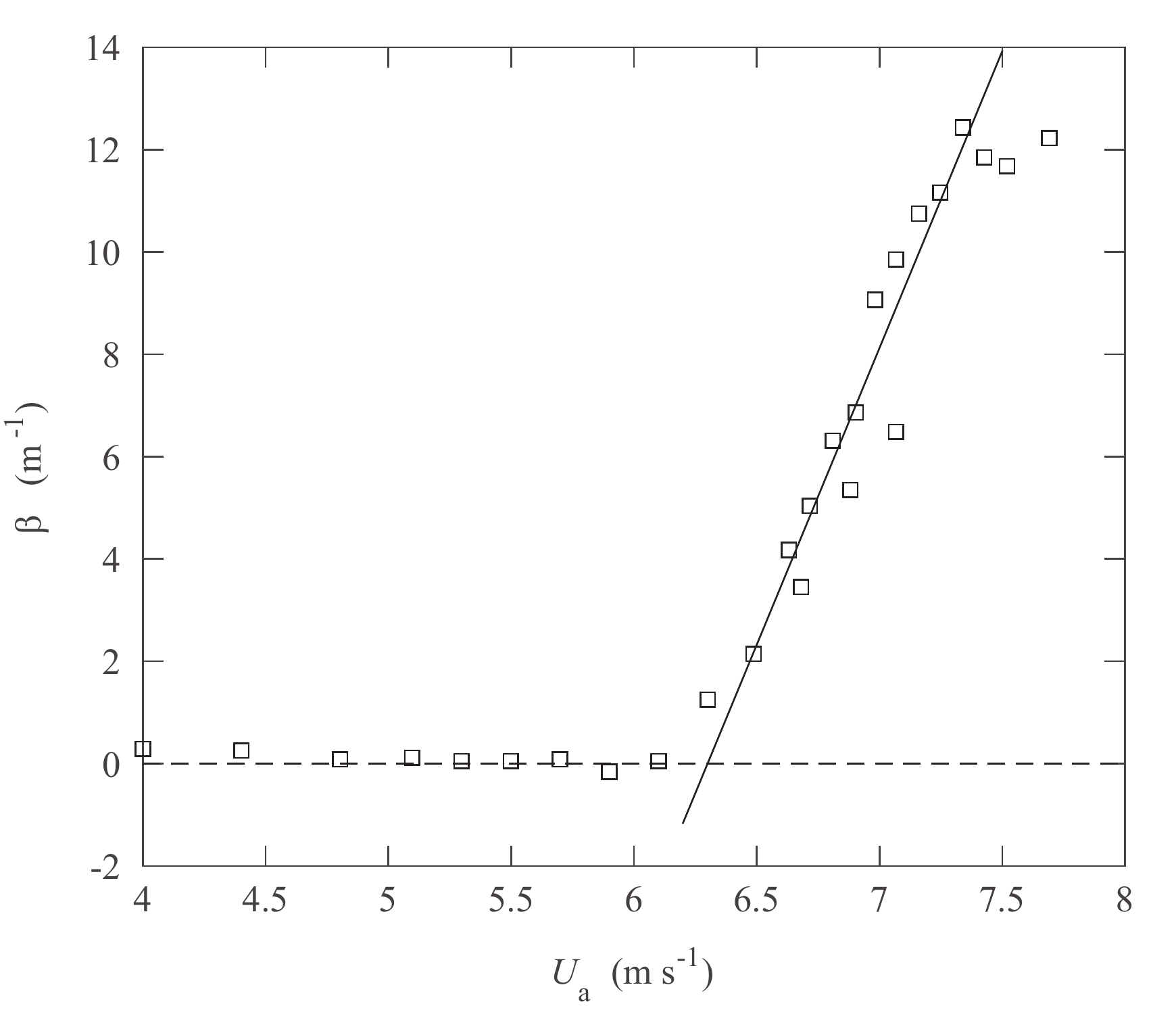}}
\caption{Spatial growth rate $\beta$, measured at small fetch ($x < 400$~mm), as
a function of wind speed. The continuous line corresponds to the linear fit (\ref{eq:fitbeta}).}
\label{fig:beta}
\end{figure}

The spatial growth rate $\beta$ can be estimated in the initial exponential growth regime ($x < 400$~mm) by fitting the squared amplitude as $\zeta^2_{\rm rms}(x) \propto \exp(\beta x)$. The growth rate $\beta$, plotted in Fig.~\ref{fig:beta} as a function of the wind velocity, allows to accurately define the onset of the wave growth: one has $\beta \simeq 0$ for $U_a < 6.3$~\mps, and a linear increase at larger $U_a$, which can be fitted by
\begin{equation}
\beta \simeq b (U_a-U_{c})
\label{eq:fitbeta}
\end{equation}
with $U_{c} \simeq 6.3 \pm 0.1$~\mps~ and $b \simeq 11.6 \pm 0.8$~s~m$^{-2}$. \modif{Interestingly, in the small range of fetch where $\beta$ is computed, the waves are nearly monochromatic, with $\lambda \simeq 30$~mm (see Sec.~\ref{sec:Spatial structures}). This indicates that the growth rate measured here, although computed from the total wave amplitude, corresponds essentially to the growth rate of the most unstable wavelength.}

\modif{We note that  the velocity threshold  $U_c\simeq 6.3$~\mps~ turns out to be close to the (inviscid) Kelvin-Helmholtz prediction. This agreement, first noted by Francis \cite{Francis_1954} for a viscous fluid, is however coincidental since the threshold depends on viscosity.\cite{keulegan1951wind}}

\modif{An interesting question is whether the wrinkles at low wind velocity can be considered as the seed noise for the exponential growth of the waves at larger velocity. Figure~\ref{fig:rmsvsfetch} indicates that this is apparently not the case:} for $U_a > 6.3$~\mps, the initial wave amplitude $\zeta_{\rm rms}$ extrapolated at $x=0$ increases with $U_a$ much more rapidly than the amplitude of the wrinkles; $\zeta_{\rm rms}(x=0)$ grows from 8 to $30 \mu$m for $U_a$ increasing from 6.3 to 7.7~\mps~ only. This suggests that the wrinkles are not necessary for the growth of the waves.
Instead, the seed noise for the waves probably results from the surface disturbance induced by the sudden change in the boundary condition from no-slip to free-slip at $x = 0$.  Accordingly, the rms amplitude can be described as the sum of the wrinkle amplitude (linearly increasing with $U_a$) and the wave amplitude (exponentially increasing with $U_a$),
\begin{equation}
\zeta_{\rm rms} \simeq \alpha U_a + \zeta_n(U_a)\exp \left[b\left(U_a-U_c\right)x\right],
\label{eq:fitadditif}
\end{equation}
with $\zeta_n(U_a)$ the amplitude of the noise at zero fetch.  

%%%%%%%%%%%%%%
%%%%%%%%%%%%%%

To provide comparison with other experiments and theoretical results, it is interesting to express Eq.~(\ref{eq:fitbeta}) in terms of a temporal growth rate.
In the frame moving with the group velocity, this temporal growth rate writes $\beta_t = c_g \beta$, with $c_g \simeq 0.16$~\mps~ for the most unstable wavelength $\lambda \simeq 30$~mm.  In terms of the friction velocity $u^* \simeq 0.05 U_a$, Eq.~(\ref{eq:fitbeta}) writes
$$
\beta_t \simeq (36 \pm 8) (u^* - 0.31)
$$
(in $s^{-1}$). Not surprisingly, these values are smaller than the ones reported in air-water experiments,\cite{Miles1959_part2, plant1982relationship, mitsuyasu1982wind, larson1975wind, snyder1966field} by a factor of order of 3, suggesting that the wave growth is weakened by the viscosity of the liquid.

\subsection{Spatial structures}
 \label{sec:Spatial structures}
 
\begin{figure}
\centerline{\includegraphics[width=8cm]{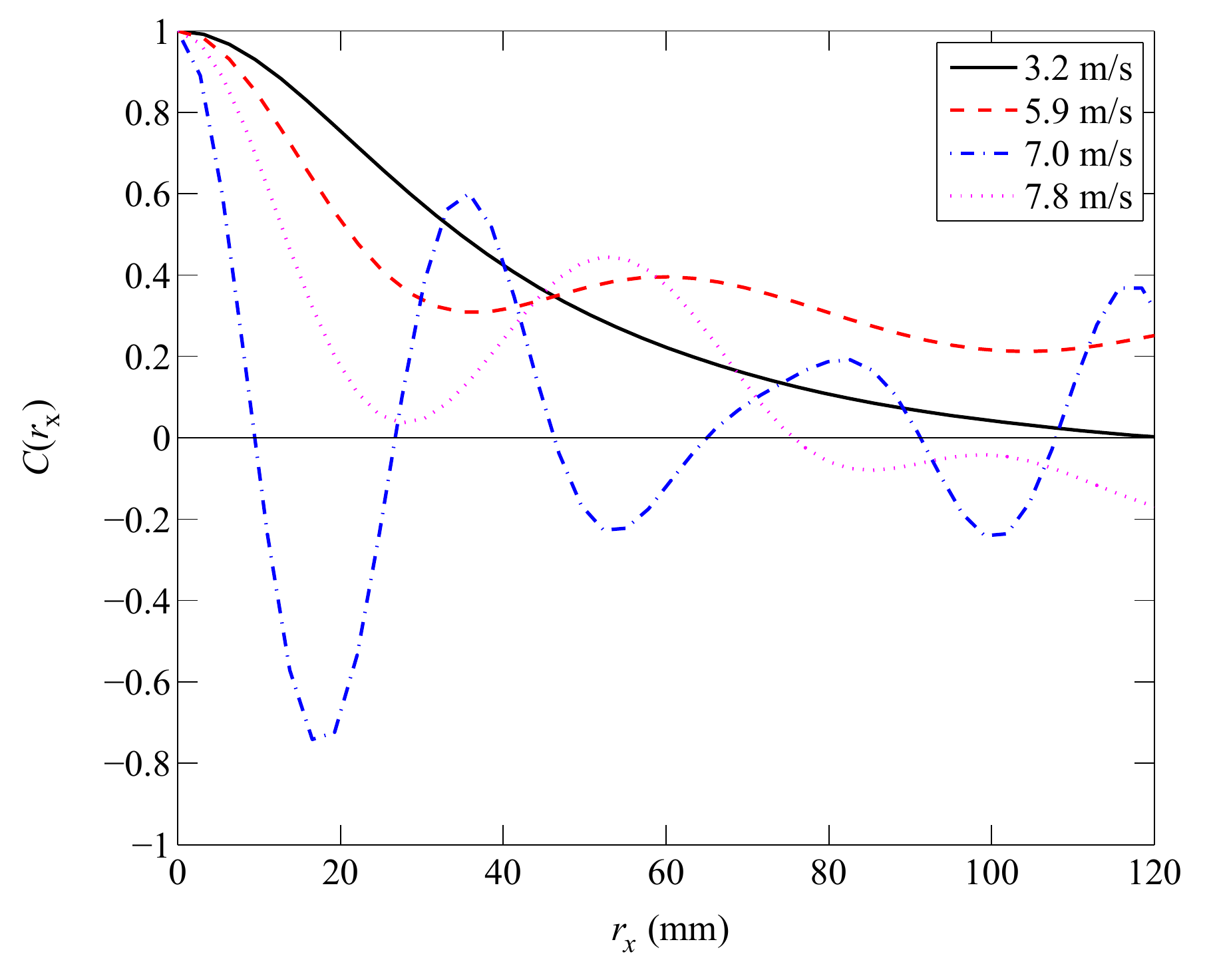}}
\caption{Spatial correlation function in the streamwise direction, $C(r_x)$, at four wind velocities for $x \in [370, 760]$~mm.}
\label{fig:correl}
\end{figure}

To characterize the spatial structure of the wrinkles and waves, we introduce the two-point correlation
$$
C({\bf r}) = \frac{\langle \zeta({\bf x},t) \zeta({\bf x} + {\bf r},t) \rangle}{\langle \zeta(x,y,t)^2 \rangle},
$$
where $\langle \cdot \rangle$ is a spatial and temporal average. The correlation in the streamwise direction (${\bf r} = r_x {\bf e}_x$) is plotted in Fig.~\ref{fig:correl} for the four wind velocities corresponding to the snapshots in Fig.~\ref{fig:snapshots}. The monotonic decay of $C(r_x)$ at small wind velocity is a signature of the disordered deformation pattern in the wrinkle regime, 
whereas the oscillations at larger velocity characterize the onset of waves. Interestingly, these oscillations are clearly visible even at $U_a = 5.9$~\mps, confirming that the transverse waves are already present in the deformation field significantly before the critical velocity $U_c \simeq 6.3$~\mps (see Fig.~\ref{fig:snapshots}b). This indicates that the transition between the wrinkles and the waves is not sharp: both structures can be found with different relative amplitude over a significant range of wind velocity.

%%%%%%%%%%%%%%
%%%%%%%%%%%%%%

The smooth transition between wrinkles and waves can be further characterized by computing the correlation length $\Lambda_i$ in the direction ${\bf e}_i$ ($i=x,y$), which we define as 6 times the first value of $r_i$ satisfying $C(r_i) = 1/2$. This definition is chosen so that $\Lambda_i$ coincides with the wavelength for a monochromatic wave propagating in the direction ${\bf e}_i$. Although no wavelength can be defined for the disorganized wrinkles, $\Lambda_x$ and $\Lambda_y$ provide estimates for the characteristic distance between wrinkles in the streamwise and spanwise directions.

\begin{figure}
\centerline{\includegraphics[width=10cm]{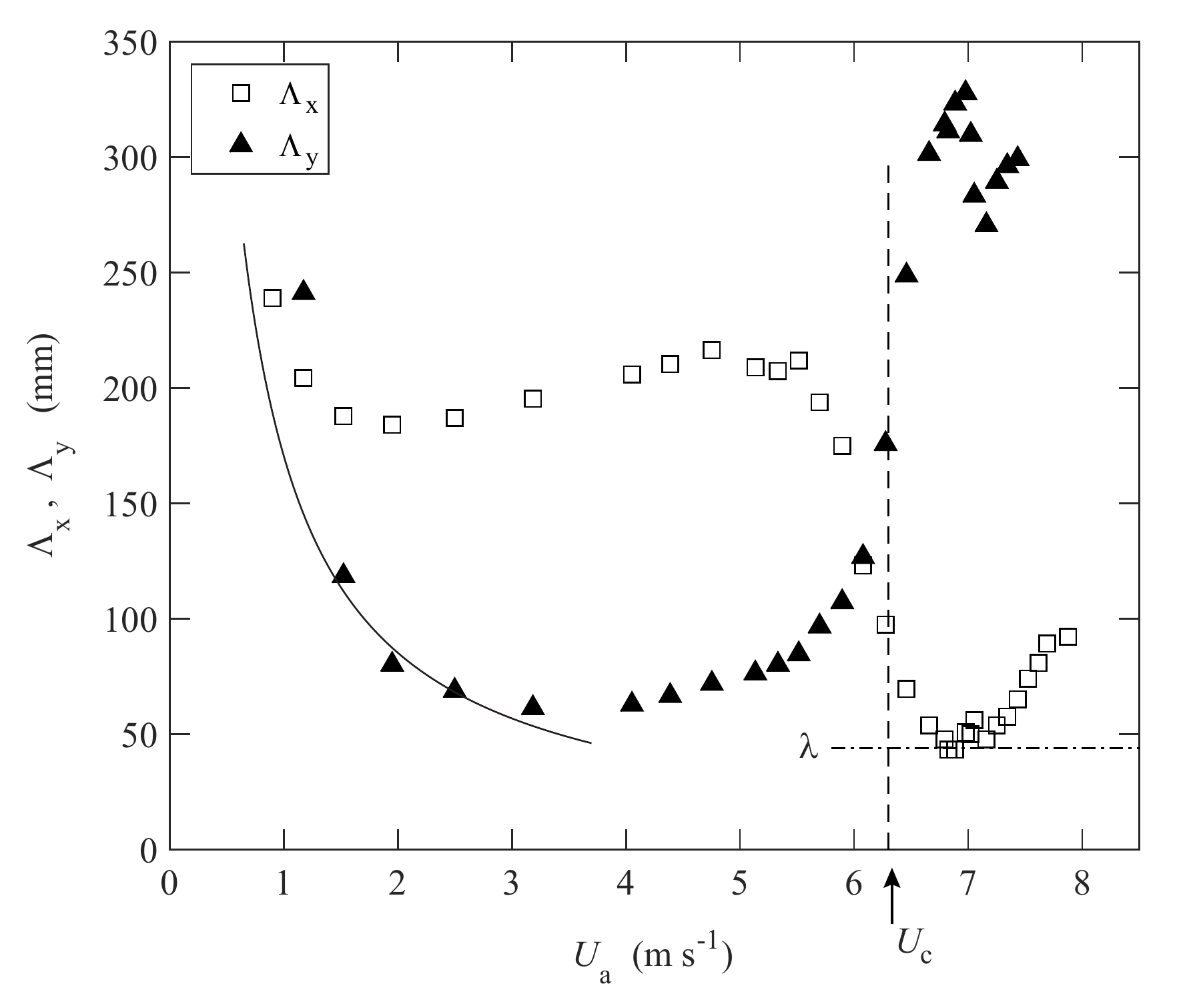}}
\caption{Streamwise and spanwise correlation lengths $\Lambda_x$ ($\square$) and $\Lambda_y$ ($\blacktriangle$), averaged over $x \in [700, 1090]$~mm, as a function of the wind velocity $U_a$. The vertical dashed line at $U_a = 6.3$~\mps~ indicates the onset of the wave growth. The continuous line is a fit $\Lambda_y \simeq 550 \delta_v$, with $\delta_v = \nu_a / u^*$ the thickness of the viscous sublayer.}
\label{fig:Lxy}
\end{figure}

The correlation lengths $\Lambda_x$ and $\Lambda_y$ are shown in Fig.~\ref{fig:Lxy} as functions of the wind velocity.  At very low wind ($\sim 1$~\mps), both lengths are of the same order, $\Lambda_x \simeq \Lambda_y \simeq 250$~mm. Between 1.5 and 5~\mps, the surface deformations are mostly in the streamwise direction ($\Lambda_x /\Lambda_y \simeq 3$), whereas at larger velocity they are essentially in the spanwise direction ($\Lambda_x /\Lambda_y \simeq 0.15$). Note that pure monochromatic waves in the $x$ direction would yield $C(r_y)=1$ and hence $\Lambda_y = \infty$; the saturation of $\Lambda_y$ close to the channel width ($W=296$~mm) observed at large $U_a$ is a signature of the dislocation existing near the center line $y=0$ and visible in Fig.~\ref{fig:snapshots}(c,d).

It is worth noting that the increase of $\Lambda_y$ and the decrease of $\Lambda_x$ start at a wind velocity $U_a \simeq 5$~\mps~ which is significantly lower than the critical velocity $U_c \simeq 6.3$~\mps, confirming that transverse waves are present well before their amplification threshold. This coexistence of wrinkles and waves at $U_a < U_c$ suggests the following picture: below the onset, waves are locally excited by the wrinkles, but they are exponentially damped ($\beta < 0$). The surface field can therefore be described as the sum of a large number of spatially decaying transverse waves, locally excited by the randomly distributed wrinkles generated by the pressure fluctuations in the boundary layer. Since the amplitude of the wrinkles is essentially independent of $x$, the resulting mixture of wrinkles and decaying transverse waves is also independent of $x$, leading to the apparent growth rate $\beta=0$ of Fig. \ref{fig:beta}. In other words, the expected negative growth rate below the onset is hidden by the spatial average over randomly distributed decaying waves, and cannot be inferred from the observed constant deformation amplitude for $U_a < U_c$.

If the elongated wrinkles at low velocity are traces of the pressure fluctuations in the boundary layers, we expect a relationship between their characteristic dimensions. The geometrical and statistical properties of the pressure fluctuations in a turbulent channel 
cannot be obtained experimentally, but are available from numerical simulations. We refer here to data from Jimenez and Hoyas\cite{jimenez2008turbulent} at $Re_\tau$ up to 2000. The intensity of the pressure fluctuations increases logarithmically from the center of the channel down to $z \simeq 30 \delta_v$, and then remains essentially constant in the viscous sublayer down to $z=0$ (see their Fig.~8b). In the thin region where the pressure fluctuations are maximum, $0<z<30 \delta_v$, the characteristic dimensions of the pressure structures in the $(x,y)$ planes normal to the wall are nearly equal, $\ell_x \simeq \ell_y \simeq 160 \delta_v$, and are hence decreasing with increasing wind velocity. These features are indeed compatible with the correlation lengths in Fig.~\ref{fig:Lxy}: at very low velocity the wrinkles are nearly isotropic, $\Lambda_x \simeq \Lambda_y \simeq 250$~mm. As $U_a$ increases up to 3~\mps, the spanwise correlation length decreases as $\Lambda_y \simeq 550 \delta_v$, similarly to the width $\ell_y$ of the pressure fluctuations. This decrease is however not observed for the streamwise correlation length $\Lambda_x$, which may result from the viscous time response of the surface behind a moving pressure perturbation.

%%%%%%%%%%%%%%
%%%%%%%%%%%%%%

\subsection{Spatiotemporal dynamics}

\begin{figure}
\centerline{\includegraphics[width=12cm]{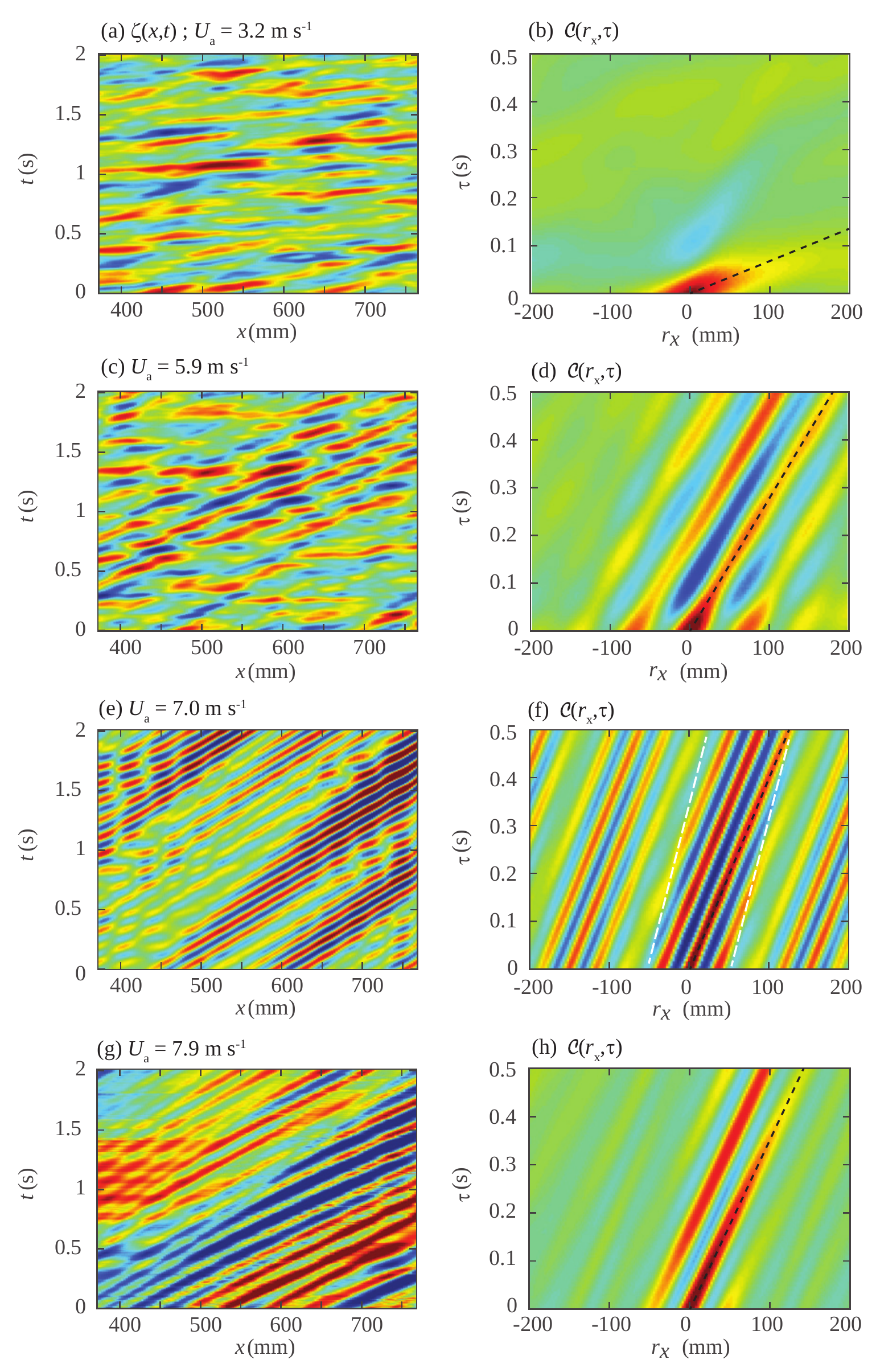}}
\caption{(a,c,e,g) Spatio-temporal diagrams $\zeta(x,t)$ taken along the line $y=0$. Same velocities and scales as in Fig.~\ref{fig:snapshots}.
(b,d,f,h) Longitudinal spatio-temporal correlation ${\cal C}(r_x,\tau)$ [Eq.~(\ref{eq:crt})]. Colormap is $[-1,1]$ from blue to red. The black dashed line shows the convection velocity $V_{\rm conv}$ (\ref{eq:vcorr}). In (f), the two white dashed lines show the group velocity $c_g = 0.16$~\mps~ corresponding to the observed wavelength $\lambda=37$~mm, which delimit the wave packets.}
\label{fig:combo}
\end{figure}

In order to confirm the relation between the surface wrinkles and the pressure fluctuations traveling in the boundary layer, we now turn to a spatio-temporal description of the surface deformation. We show in Fig.~\ref{fig:combo} spatio-temporal diagrams (left)
and two-point two-time correlation (right) at increasing wind velocity. The spatio-temporal diagrams are constructed by plotting the surface deformation $\zeta(x,y,t)$ in the plane $(x,t)$ along the center line $y=0$. The oblique lines in these diagrams indicate the characteristic velocity of the deformation patterns. The spatio-temporal correlation in the streamwise direction is defined as
\begin{equation}
{\cal C}(r_x,\tau) = \frac{\langle \zeta({\bf x} + r_x {\bf e}_x,t+\tau) \zeta({\bf x},t) \rangle}{\langle \zeta({\bf x},t)^2 \rangle},
\label{eq:crt}
\end{equation}
where $\langle \cdot \rangle$ is a spatial and temporal average. 
For statistically stationary and homogeneous deformations, one has ${\cal C}(-r_x,-\tau) = {\cal C}(r_x,\tau)$, so only the positive time domain is shown. 

%%%%%%%%%%%%%%
%%%%%%%%%%%%%%

At small wind velocity ($U_a = 3.2$~\mps) the surface deformation shows rapidly propagating disorganized structures, with life time of order of their transit time [Fig.~\ref{fig:combo}(a)]. Their characteristic velocity is distributed over a large range, resulting in a broad correlation in Fig.~\ref{fig:combo}(b).

In the mixed wrinkle-wave regime ($U_a = 5.9$~\mps), slow wave packets with well defined velocity appear, embedded in a sea of rapid disorganized fluctuations [Fig.~\ref{fig:combo}(c)]. These slow wave packets confirm the picture of transverse waves locally excited by the wrinkles but rapidly damped because of their negative growth rate. The wavelength and the phase velocity of these evanescent  waves can be inferred from the corresponding spatio-temporal correlation [Fig.~\ref{fig:combo}(d)].

At larger wind velocity ($U_a = 7.0$~\mps), the surface deformation becomes dominated by spatially growing transverse waves ($\beta>0$), resulting in well defined oblique lines in  the spatio-temporal diagram [Fig.~\ref{fig:combo}(e)] and a marked spatial and temporal periodicity in the correlation [Fig.~\ref{fig:combo}(f)]. These transverse waves, however, are never strictly monochromatic: wave packets are still visible, delimited by boundaries propagating at the group velocity.  For this wind velocity and fetch, the local wavelength is $\lambda = 37$~mm, for which the predicted phase and group velocities are $c=0.26$~\mps~ and $c_g=0.16$~\mps, in good agreement with the observed slopes (black and white dashed lines, respectively) in Fig.~\ref{fig:combo}(f).

Finally, at even larger wind velocity ($U_a=7.9$~\mps) the phase velocity and wavelength increase slightly, and becomes again broadly distributed. Accordingly, the spatial and temporal periodicity weakens in Fig.~\ref{fig:combo}(h).

%%%%%%%%%%%%%%
\begin{figure}
\centerline{\includegraphics[width=10cm]{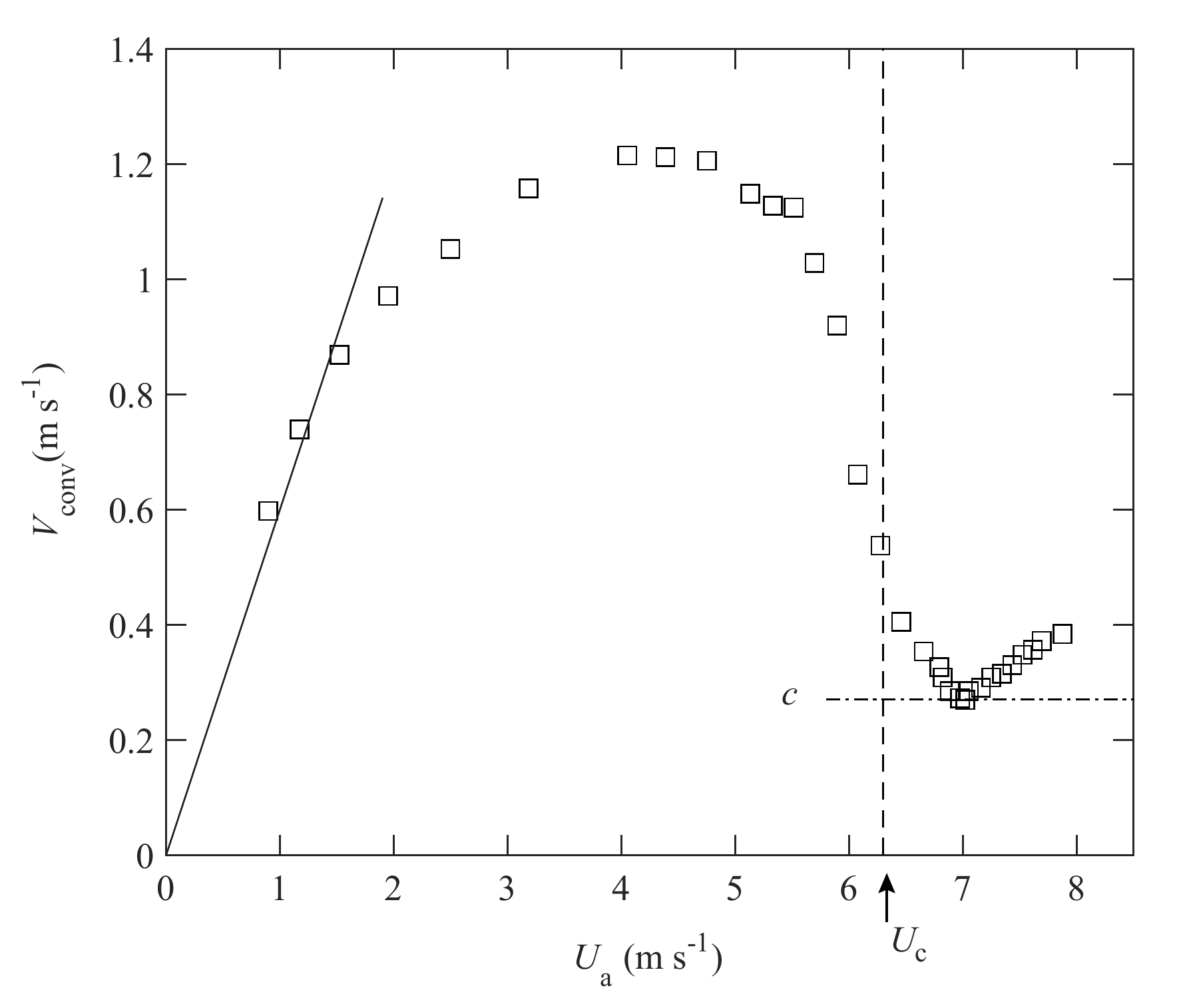}}
\caption{Convection velocity $V_{\rm conv} = \Lambda_x / \tau$, averaged over $x \in [700, 1090]$~mm, as a function of the wind velocity $U_a$. The vertical dashed line at $U_c = 6.3$~\mps~ indicates the onset of wave growth, $\beta=0$. The continuous line shows $V_{\rm corr} = 0.6 U_a$. The dashed horizontal line shows the local phase velocity $c=0.27$~\mps~ corresponding to the local wavelength $\lambda = 44$~mm.}
\label{fig:Vcorr}
\end{figure}
%%%%%%%%%%%%%%

For monochromatic waves propagating in the $x$ direction, one has ${\cal C}(r_x,\tau) = \cos (k r_x - \omega t)$, so the correlation is 1 along characteristic lines parallel to $r_x / t = c$, where $c = \omega/k$ is the phase velocity. For a (non-monochromatic) propagating pattern, the correlation is weaker but remains maximum along a line $r_x/t$ given by the characteristic velocity of the pattern. We therefore define the \modif{convection velocity} as
\begin{equation}
V_{\rm conv} = \frac{\Lambda_x}{\tau},
\label{eq:vcorr}
\end{equation}
where $\Lambda_x$ and $\tau$ are defined as the first value such that ${\cal C}(\Lambda_x/6,0)=1/2$ and ${\cal C}(0,\tau/6)=1/2$. Here again, the factor 6 is chosen such that $\Lambda_x$ and $\tau$ correspond to the wavelength and period for monochromatic waves; in this case, $V_{\rm conv}$ is simply the phase velocity. This convection velocity is shown as black dashed lines in the spatio-temporal correlation in Fig.~\ref{fig:combo}, and is plotted in Fig.~\ref{fig:Vcorr} as a function of the wind velocity $U_a$. At very small velocity ($U_a < 2$~\mps), $V_{\rm conv}$ first increases, following approximately $V_{\rm conv} \simeq 0.6 U_a$. \modif{This value is remarkably similar to the convection velocity of the pressure fluctuations found in turbulent boundary layers.\cite{choi1990space} This is consistent with the fact that, in this small range of $U_a$, the decrease of the characteristic width of the wrinkles $\Lambda_y$ follows the expected decrease of the size of the pressure fluctuations in the boundary layer (Fig.~\ref{fig:Lxy}).   Such convection velocity $V_{\rm conv} \simeq 0.6 U_a$ can be recovered as follows: in a turbulent boundary layer} the pressure fluctuations are maximum at $z_m / \delta_v \simeq 20-50$ (see, e.g., Jimenez and Hoyas\cite{jimenez2008turbulent}) and, according to the logarithmic law (\ref{eq:CL_Turbulente}), the mean velocity at this height is $u(z_m) \simeq (13\pm 1) u^*$. Using $u^* \simeq 0.05 U_a$ in the present experiment, this yields $u(z_m) \simeq (0.65 \pm 0.05) U_a$. This suggests that the surface response to the traveling pressure fluctuations is essentially local and instantaneous up to $U_a \simeq 2$~\mps.

\modif{For $U_a > 2$~\mps, the convection velocity departs from the linear growth $0.6 U_a$, and saturates at $V_{\rm conv} \simeq 1.2$~\mps. This saturation probably results from the viscous damping, which prevents an instantaneous response of the surface deformation to too rapidly propagating pressure disturbances. 
As the wind velocity is further increased, the convection velocity decreases down to $c \simeq 0.27$~\mps, which coincides with the expected phase velocity of free waves of the observed wavelength}. This gradual decrease does not correspond to a slowing of the wrinkles, but rather to an average over the rapid wrinkles propagating at velocities of the order of $1.2$~\mps, which dominate the surface at low $U_a$, and the slow transverse waves at velocity of the order of $0.27$~\mps, which dominate the surface at high $U_a$. \modif{Finally, for $U_a > 7$~\mps, the convection velocity increases again, which is consistent with the increase of the wavelength in Fig.~\ref{fig:Lxy}.
These increases do not necessarily occur for shorter fetches, implying that the wave properties change with fetch at high wind velocity. Such non linear effect presents strong similarities with the wavenumber and frequency downshift observed for wind-waves on sea, and will be investigated in future work."}

\section{Conclusion}
\label{sec:conclusion}

In this paper, we explored the spatio-temporal properties of the first surface deformations induced by a turbulent wind on a viscous fluid. New insight into the wave generation mechanism is gained from spatio-temporal correlations computed from high resolution time-resolved measurements of the surface deformation field. At low wind velocity, rapidly propagating disordered wrinkles of very small amplitude are observed, resulting from the response of the surface to the traveling pressure fluctuations in the turbulent boundary layer. Above a critical wind velocity $U_c$, we observe the growth of well defined propagating waves, with growth rates compatible with a convective supercritical instability. Interestingly, an intermediate regime with spatially damped waves locally excited by the wrinkles is observed below $U_c$, resulting in a smooth evolution of the characteristic lengths and velocity as the wind speed is increased. Above the onset $U_c$, the seed noise for the growth of the waves is apparently not governed by the wrinkles, but rather by the perturbations at the inlet boundary condition at zero fetch.

Using a liquid of large viscosity yields considerable simplification of the general problem of wave generation by wind. Although some of the present results may be relevant to the more complex air-water configuration, other are certainly specific to the large viscosity of the liquid. In particular, the wrinkles observed at very low wind velocity ($U_a < 2$~\mps) are compatible with a local and instantaneous response of the surface to the pressure fluctuations traveling in the boundary layer. This simple property is not expected to hold for liquids of lower viscosity such as water, for which the surface deformation at a given point results from the superposition of the disturbances emitted previously from all the surface. New experiments with varying viscosity are necessary to gain better insight into the intricate relation between the turbulent pressure field and the surface response below the onset of the wave growth, and to characterize the spatial evolution of the waves above onset.

\acknowledgments

We are grateful to H. Branger, C. Clanet, P. Clark di Leoni, B. Gallet, J. Jim\'enez, \'O~N\'araigh, and P. Spelt for fruitful discussions. We acknowledge A. Aubertin, L. Auffray, C. Borget, and R. Pidoux for the design and set-up of the experiment. This work is supported by RTRA "Triangle de la Physique". F.M. thanks the Institut Universitaire de France.

%%%%%%%%%%%%%
\bibliographystyle{unsrt}
\bibliography{biblioWindWaves}
%%%%%%%

\end{document}